\documentclass[aps,prl,twocolumn,groupedaddress,showpacs]{revtex4}
\usepackage{dcolumn}
\usepackage{epsfig}
\usepackage{graphicx}
\usepackage{color}

\begin{document}
%%%%%%%%%%%%%%%%%%%%     TITLE    %%%%%%%%%%%%%%%%%%%%%%%%%%%%%
\title{NiTi shape-memory transformations: minimum-energy pathways between austenite, martensites, and kinetically-limited intermediate states}
\author{N.~A. Zarkevich$^{1}$ and D.~D. Johnson$^{1,2}$}
\email{zarkev@ameslab.gov, ddj@ameslab.gov}
\affiliation{$^{1}$The Ames Laboratory, U.S. Department of Energy, Ames, Iowa 50011-3020 USA;}
\affiliation{$^{2}$Materials Science \& Engineering, Iowa State University, Ames, Iowa 50011-2300 USA.}

\date{\today}
\begin{abstract} 
NiTi is the most used shape-memory alloy, nonetheless, a lack of understanding remains regarding the associated structures and transitions, including their barriers. 
Using a generalized solid-state nudge elastic band (GSSNEB) method implemented via density-functional theory, we detail the structural transformations in NiTi relevant to shape memory: those between body-centered orthorhombic (BCO) groundstate and a newly identified stable austenite (``glassy'' B2-like) structure, including energy barriers (hysteresis) and intermediate structures (observed as a kinetically limited R-phase), and between  martensite variants (BCO orientations). All results are in good agreement with available experiment. We contrast the austenite results to those from the often-assumed, but unstable B2. These high- and low-temperature structures and structural transformations provide much needed atomic-scale detail for transitions responsible for NiTi shape-memory effects.
\end{abstract}
\maketitle

\begin{figure}[b]
\includegraphics[width=80mm]{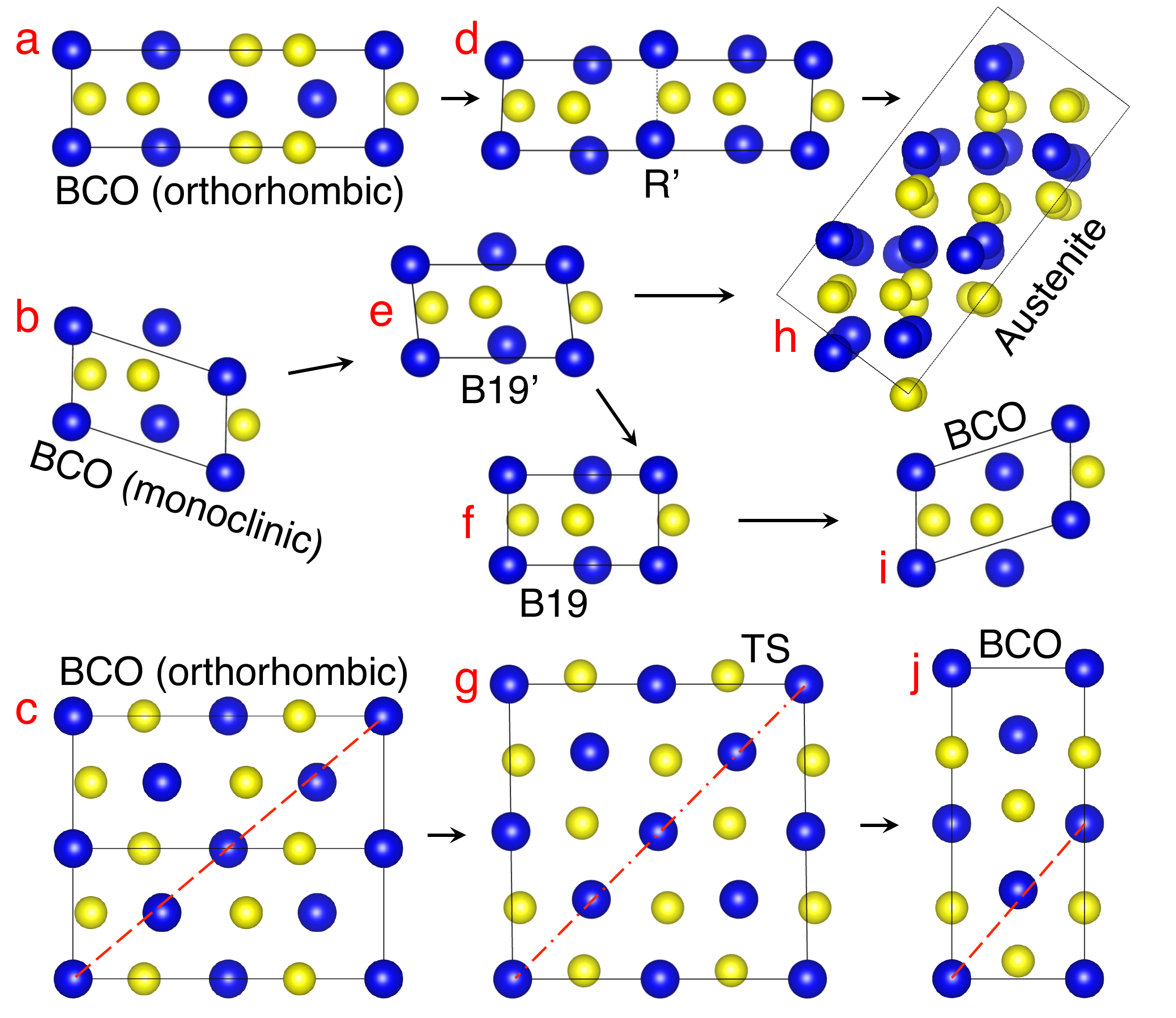}
\vspace{-2mm}
\caption{\label{Fig2path}(color online) 
Relevant structures [Ni (smaller, yellow) and Ti (larger, blue)] in $[010]_{\text{BCO}}$ (a, b, d, e, f, h, i) and $[001]_{\text{BCO}}$ (c, g, j) projections. 
BCO (a, b, c, i, j), Austenite (h), and intermediates: R' (d) and B19' (e).
Shown also are BCO-to-BCO transition states: B19 (f) and twin (g). 
}
\end{figure}

%\section{\label{sIntroduction}Introduction}
{\par } NiTi (near 50 at.\% Ti), or Nitinol \cite{JAP34p475y1963} is the most used shape-memory alloy, recovering its original shape upon heating even after a substantial mechanical deformation \cite{SMA2006}.   
It finds applications in medical implants, industrial devices, 
thermally activated robotics at nano- and macro-scales, reading-glass frames, and heated reshaping of brassieres -- its most profitable use. 
In spite of such intensive use, the underlying physics and atomistic mechanics of the shape-memory effect in this ``simple'' binary alloy remain unclear. 
First and foremost, rather than a simple (but unstable) B2 structure  always used heretofore, the stable high-temperature austenite phase was recently discovered to be a more complex structure \cite{PRB90p060102R}, with configurations displaying prominent, correlated static displacements but behaves B2-like on average, acting similar to a phonon glass \cite{RevModPhys86p669y2014}. 
Here, using the GSSNEB method \cite{GSSNEB} implemented via density-functional theory (DFT), we consider the NiTi shape-memory transformations (without use of intuition) that involve the proposed NiTi groundstate base-centered orthorhombic (BCO) structure \cite{nmat2p307y2003} and recently discovered stable austenite structure \cite{PRB90p060102R}, along with observed B19' and R phases, see Fig.~\ref{Fig2path}.

%%%%%%%%%%%%%%%%%%%%%%%%%%%%%%%%%%%%%%%%%%%%%%%%%%%%%%%%
%%%%%%%%%%%%%%%%%%%%%%%%%%%%%%%%%%%%%%%%%%%%%%%%%%%%%%%%

\begin{figure}[t]
\includegraphics[width=80mm]{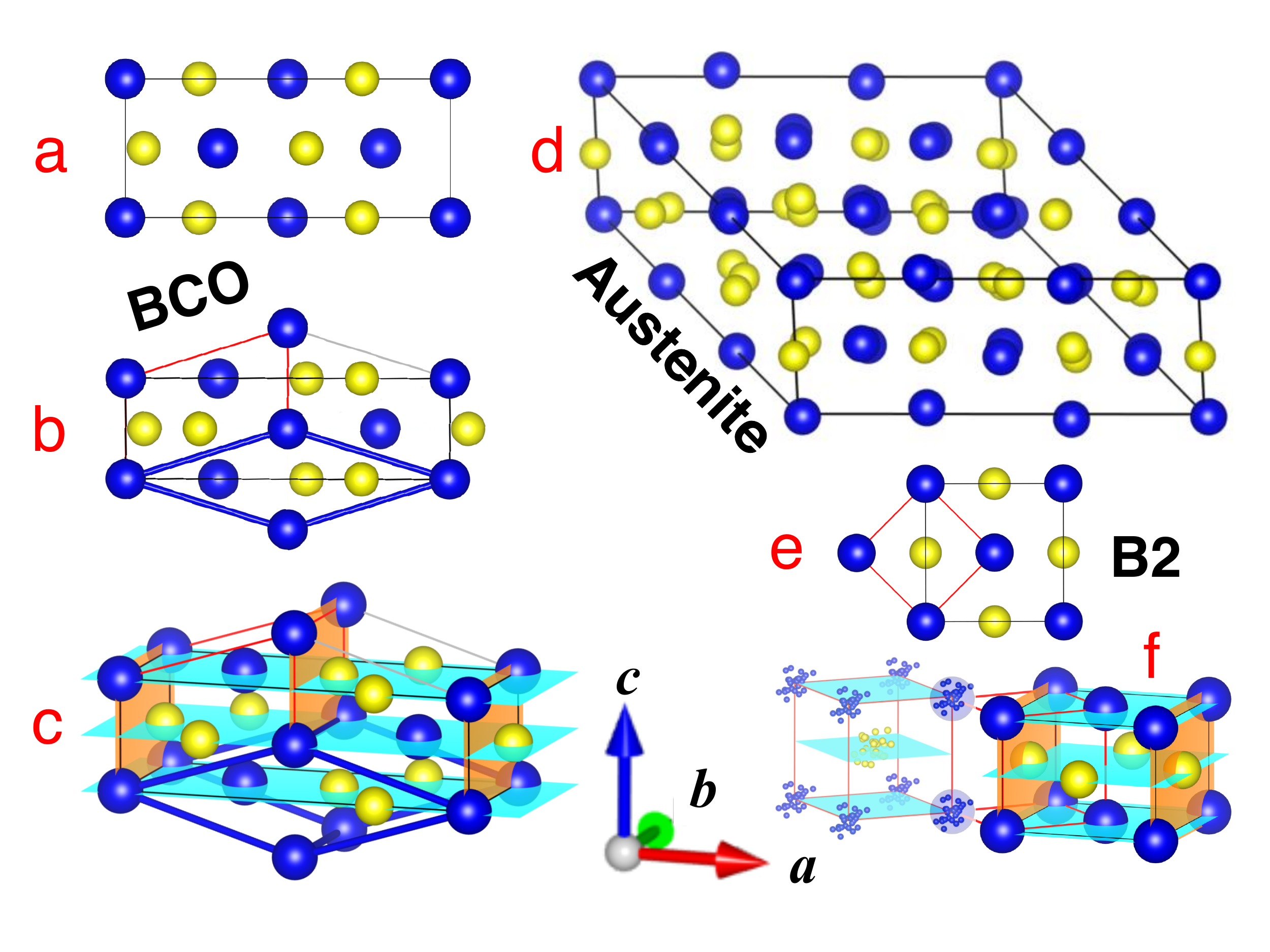}
\caption{\label{Fig1struc}(color online) 
(a) $[001]$ and (b) $[010]$ projections of (c) BCO;
(d) stable austenite (Fig.~\ref{Fig2path}h, represented by 54-atom hexagonal cell [\onlinecite{PRB90p060102R}]), compared to (e) $[001]$ projection of (f) unstable B2 (see left-side overlay of austenite displacements projected onto B2).}
\end{figure}

{\par} To understand shape-memory transformations, the atomic structures of the endpoint martensites and austenite are required. 
The NiTi groundstate is accepted as BCO \cite{nmat2p307y2003,PRB90p060102R,GudaVishnu2010745,PRB85p014114}. 
The smallest stable austenite structure has a hexagonal (Ni$_{27}$Ti$_{27}$)  cell \cite{PRB90p060102R}, Fig.~\ref{Fig1struc}, and DFT yields agree well with all available calorimetry, x-ray diffraction (structure) and neutron diffraction (phonon density of states) data \cite{PRB90p060102R}. 
Multiple attempts have been made to visualize via atomic-scale simulations the NiTi austenite-to-martensite transformation (Fig.~\ref{Fig2path}), all with B2-austenite assumed.
For example, molecular dynamics simulations \cite{MaterTrans47p742y2006,ActaMat60p6301,JAP110p033532y2011} based on a semi-empirical potentials yielded two major stress-induced transformation from B2(austenite)-to-B19'(martensite), missing the observed BCO groundstate with B19' lower than BCO by $8~me$V/NiTi \cite{JAP110p033532y2011}.  
Shear-shuffle models of detwinning were considered using DFT \cite{ActaMat60p339y2012,MatSciEngA558p442y2012,ActaMat59p5893y2011,APL98p241906,APL98p141906,ActaMat61p67y2013,B2ddj9},  some guided by intuition, yet there has been no success in modeling the austenite-to-BCO martensitic transformation. 
%%%%%Below is just a list, not conclusion or relation, so Shear-shuffle above seems Sufficient%%%%
%A shear-shuffle ($\bar{2}11$)[$\bar{1}\bar{1}\bar{1}$] twinning of B2-Austenite was proposed \cite{APL98p241906}.   
%The energy of the (114)[$22\bar{1}$] twin in B2 was calculated \cite{ActaMat60p339y2012}. 
%%%  HOW IS THIS RELEVANT?  RESULT?
%A dislocation-slip deformation of B2 was simulated and compared to transmission electron microscopy \cite{ActaMat61p67y2013}.  
%Type-II detwinning in B19' was studied \cite{APL98p141906}. 
%%%  HOW IS THIS RELEVANT?  RESULT?
%The (001)[100] twin \cite{ActaMat59p5893y2011} and ($20\bar{1}$)[$\bar{1}0\bar{2}$] twin \cite{MatSciEngA558p442y2012} in B19' structure with $\vartheta$ of $97.8^\circ$ between [100] and [001] axes were calculated and compared to microscopy data  \cite{MatSciEngA558p442y2012,ActaMat59p5893y2011}.    
For a hypothetical two-step B2--B19--B19' pathway \cite{B2ddj9}, a B2--B19 transition state was $13~me$V/NiTi above B2. 
However, GSSNEB finds that the B2--BCO transformation bypasses B19 and has no barrier (reflecting the instability of B2) in agreement with experiment \cite{exptB65}. 
Hence, in spite of the considerable attention attracted by shape-memory Ni-Ti, the atomic-scale understanding of its shape memory remains highly incomplete.  

{\par} Here, we consider transformations amongst BCO variants, as well as  those between BCO, stable austenite, and unstable B2 (all shown in Fig.~\ref{Fig1struc}). 
We obtain kinetically--limited structures, similar to the observed R-phase  \cite{ProgMatSci50n5p511y2005Otsuka}.
Our calculated barrier for austenite-to-martensite compares well with the observed hysteresis width \cite{ActaMat52p3383y2004}.

%%%%%%%%%%%%%%%%%%%%%%%%%%%%%%%%%%%%%%%%%%%%%%%%%%%%%%%%
%%%%%%%%%%%%%%%%%%%%%%%%%%%%%%%%%%%%%%%%%%%%%%%%%%%%%%%%

{\par} {\bf Methods: }For the DFT-based electronic-structure method, we utilize the VASP code \cite{VASP3,VASP4}  within a generalized gradient approximation (GGA) \cite{GGA,PW91} and projected augmented wave (PAW) basis \cite{PRB50p17953}. 
We choose $337.0~e$V planewave energy cutoff and $544.6~e$V augmentation charge cutoff. Converged $k$-meshes have at least 50 $k$-points per {\AA}$^{-1}$ (e.g., $11 \times 13 \times 17$ for a $4.92 \times 4.00 \times 2.92$ {\AA} cell). We use the modified Broyden method \cite{PRB38p12807y1988} for convergence. The GSSNEB calculations are completed using Gaussian smearing with $\sigma =0.05\,e$V, and the tetrahedron method with Bl\"ochl corrections verify the energies relative to BCO.
 Two representative structures of austenite were obtained by \emph{ab initio} molecular dynamics followed by atomic relaxations at 0 K using conjugate gradient method \cite{PRB90p060102R} -- hexagonal unit cells of 54-atom and 108-atom (i.e., doubled along $a$).  The transition states (Fig.~\ref{Fig4}) are obtained using the GSSNEB method \cite{GSSNEB} modified to use double-climbing images (C2-NEB) \cite{C2NEB}, providing a more reliable minimal-energy path (MEP) for complex potential-energy landscapes. 

%%% KEEP THIS SECTION LABEL even if to PRL
\section{\label{Results}Structures and Deformations}
{\bf Ground State:} BCO structure is the DFT ground\-state \cite{nmat2p307y2003,PRB90p060102R,GudaVishnu2010745,PRB85p014114}. 
Our calculated lattice constants are $2.9217$, $4.0024$, and $4.9189$~{\AA}, with an angle of $107.23^\circ$, 
defined in the monoclinic unit cell (Fig.~\ref{Fig2path}b).
BCO consists of two interpenetrating hcp sublattices, populated by Ni and Ti, respectively, 
with slightly displaced atoms due to Ni--Ti interaction (Fig.~\ref{Fig1struc} a, b, c).

\begin{figure}[t]
%\begin{center}
\includegraphics[width=70mm]{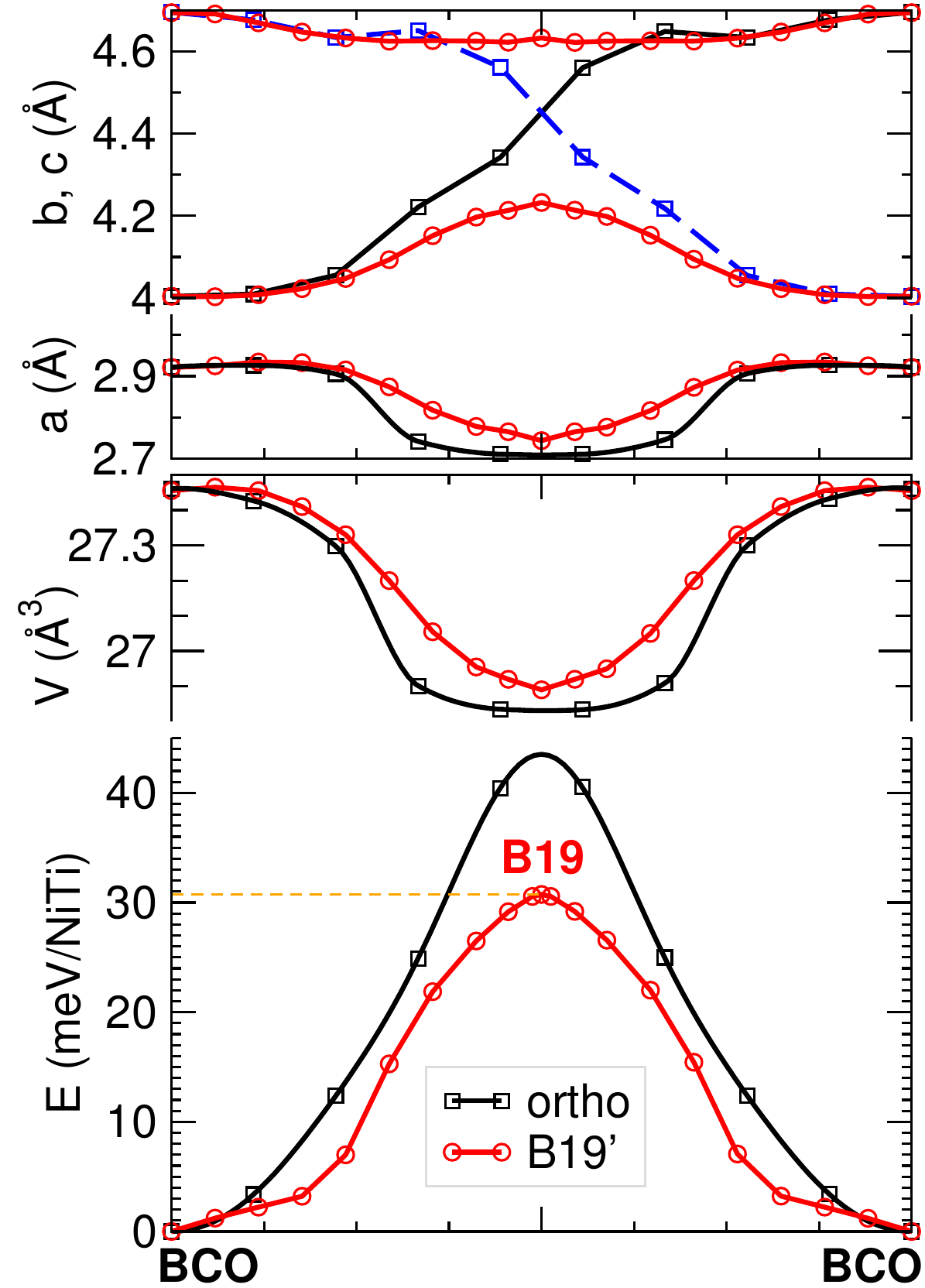}
%\end{center}
\vspace{-2mm}
\caption{\label{Fig4}(color online) 
GSSNEB MEP of BCO-to-BCO. Orthorhombic distortion (squares) interchanges the lattice constants $b$ (solid black) and $c$ (dashed blue line) . Shear of the monoclinic B19' cell (red circles) has B19 transition state;  $c \cdot \sin \theta $ is plotted as the lattice constant normal to $a$ and $b$.  
}
\end{figure}

\begin{figure}[t]
%\begin{center}
\includegraphics[width=70mm]{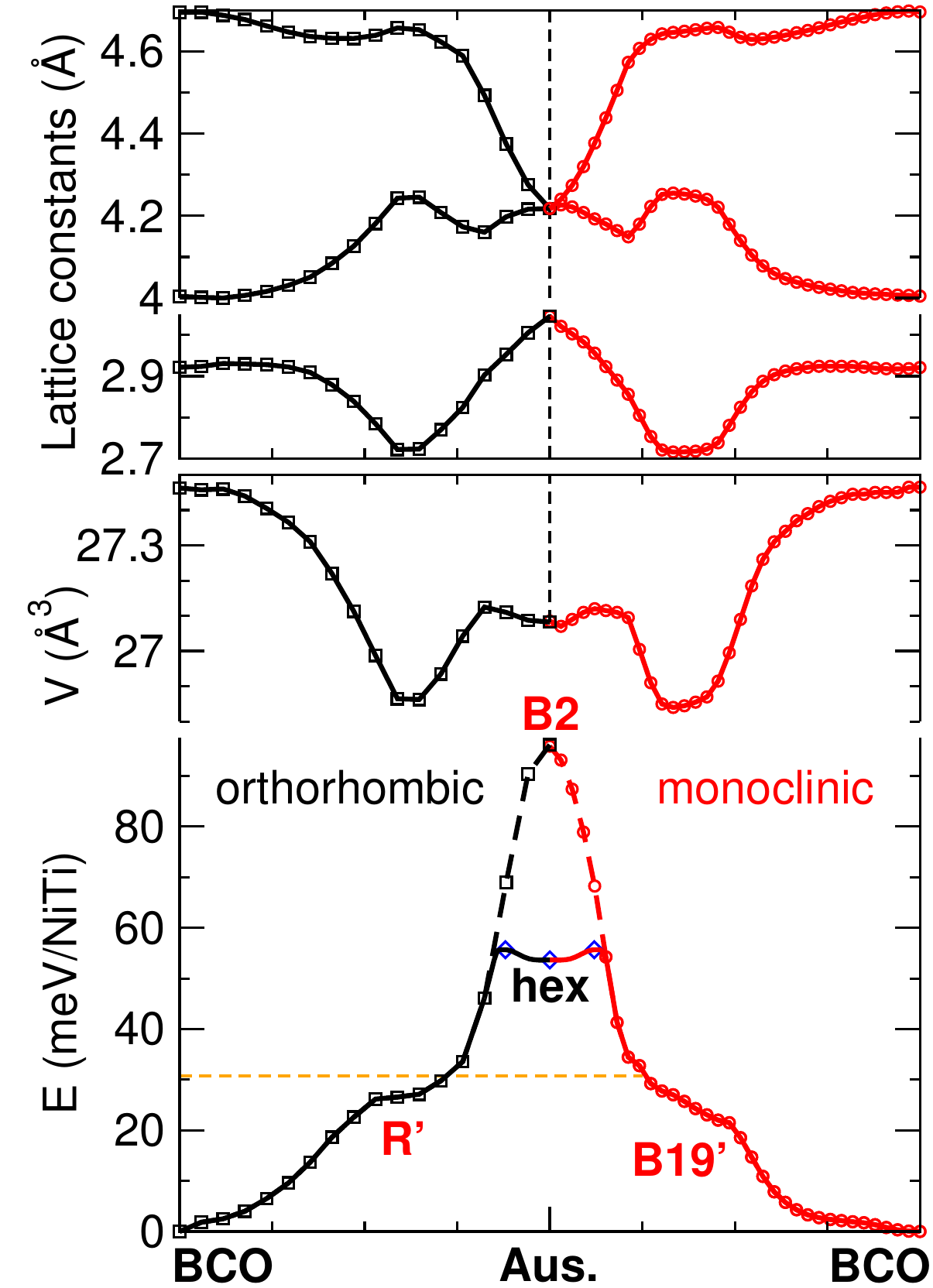}
%\end{center}
\vspace{-2mm}
\caption{\label{Fig3aus}(color online) 
GSSNEB MEP from BCO to austenite (solid line) and to unstable B2 (dashed)  
in orthorhombic (left, black) and monoclinic B19' (right, red) unit cells. 
B19 energy from Fig.~\ref{Fig4} is given by the horizontal (orange) dashed line. 
}
\end{figure}

{\bf Deformed Martensite:} Monoclinic B19' ($\vartheta \! \approx \! 98^\circ$) is a low-energy deformation of BCO (Fig.~\ref{Fig2path}e), as observed \cite{JETP77p2341y1979}. 
With BCO viewed as B19'  with $\vartheta \! \approx \! 107^\circ$, the energy of B19' is from $0$ to $16~me$V/atom higher than BCO, see Figs.~\ref{Fig4}~and~\ref{Fig3aus} (agreeing with Fig.~\ref{Fig2path}a in [\onlinecite{nmat2p307y2003}]). 
In the martensite, the experimental B19' structure is not the ground state, but its low-energy deformation, stabilized by the martensitic stress, and the ease to deform martensite accounts for its ``superelasticity''.
Perfect BCO can be  represented by B19' unit cells of two alternating or same orientations, giving boundaries between them that cost no energy (Fig.~\ref{Fig1struc}); it also produces low-energy twins. 

{\bf Deformation Twins:} Deformation of martensite is accompanied by motion of twins and other planar defects. 
Motion of twins was recently addressed in B19' and B2 structures  \cite{ActaMat60p339y2012,MatSciEngA558p442y2012,ActaMat59p5893y2011,APL98p241906,APL98p141906}.  
While pseudo-twinning in B2 NiTi has been suggested as impossible \cite{Paxton1995}, perfect B2 is unstable and its structure is not  relevant to the martensitic transformation.
Approximated by periodic twins separated by 12.2~{\AA},  DFT energy for (210)-twins in BCO is extremely low at $0.53~me$V/{\AA}$^2$ (or $8.4~m$J/m$^2$); when separated by only 6~{\AA}, the twin energy of $0.83~m$eV/{\AA}$^2$ (or $13.3~m$J/m$^2$) is higher,  so twins repel at small distances, 
in agreement with observation \cite{ProgMatSci50n5p511y2005Otsuka}.  
%observed quasi-periodic twins \cite{ProgMatSci50n5p511y2005Otsuka}.  

{\bf Austenite:} The stable austenite structure, with representative Ni$_{27}$Ti$_{27}$ hexagonal cell, has a DFT energy of $\Delta E=29.5\,$meV/atom above BCO \cite{PRB90p060102R},  giving an estimated \cite{PRB75p104203} martensitic temperature $T_c \approx \Delta E/k_B=343\,$K, near the observed \cite{MT2p229y1971} 333 K for stoichiometric NiTi. 
A larger structure for austenite-to-BCO MEP calculations is constructed from the 54-atom cell in Fig.~\ref{Fig1struc}(d) by doubling along $a$.
The resulting Ni$_{54}$Ti$_{54}$ structure (without 3~{\AA} periodicity along $a$) has the same energy and similar displacements and properties as Ni$_{27}$Ti$_{27}$ austenite \cite{PRB90p060102R}, so it too can be used to model austenite.  
This newly reported austenite looks B2 ``on average'' (Fig.~\ref{Fig1struc}f), i.e., if atomic positions are averaged into a 2-atom B2 cell \cite{PRB90p060102R}.  

{\par} For completeness, the Ni-Ti phase diagrams \cite{exptB91,ZMetallkd87p972y1996} reference the cubic B2 (Fig.~\ref{Fig1struc}~e, f) as the high-temperature solid phase.  
Our DFT B2 lattice constant is $3.0028$~{\AA }, in agreement with previous calculations  \cite{nmat2p307y2003,PRB90p060102R,GudaVishnu2010745,PRB85p014114}.  
Powder diffraction measurements \cite{exptB65,exptHsol,ZPhysChemNeueFol164p803y1989} give $3.015$~{\AA} at $353\,$K.
Our DFT  energy for B2 is 48 meV/atom above BCO, agreeing with previous calculations  \cite{nmat2p307y2003,PRB90p060102R,GudaVishnu2010745,PRB85p014114}, 
and corresponds to an estimated martensitic temperature of $557\,$K, well above that observed \cite{MT2p229y1971}. 
Importantly, B2 is known to be unstable, with imaginary phonon modes not stabilized by entropy below melting \cite{PRB90p060102R,exptB65}. 
Reflecting this instability, we find no barrier for a B2-to-BCO transition (Fig.~\ref{Fig3aus}). 

%%%%%%%%%%%%%%%%%%%%%%%%%%%%%%%%%%%%%%%%%%%%%%%%%%%%%%%%
%%%%%%%%%%%%%%%%%%%%%%%%%%%%%%%%%%%%%%%%%%%%%%%%%%%%%%%%
%%%%%%%%%%%%%%%%%%%%%%%%%%%%%%%%%%%%%%%%%%%%%%%%%%%%%%%%
%%%%%%%%%%%%%%%%%%%%%%%%%%%%%%%%%%%%%%%%%%%%%%%%%%%%%%%%

%%% KEEP THIS SECTION LABEL even if to PRL
\section{\label{Result2}MEP and Transition States}
{\par}{\bf BCO-to-BCO:} First, we examine the solid-solid transformations between various orientations of BCO (Fig.~\ref{Fig4}). 
We find several transition states, with the lowest-energy B19 being only $15.4~m$eV/atom above BCO, 
similar in structure and energy to the kinetically limited R-phase (Fig.~\ref{Fig2path}).   
Among the two TS in Fig.~\ref{Fig2path} (f, g), 
the more symmetric Ni$_8$Ti$_8$ TS (g) with the orthogonal lattice vectors $a=b \ne c$ (a=b=8.9, c=2.7 {\AA}) 
is expected to be higher in energy 
compared to B19 Ni$_2$Ti$_2$ with $a \ne b \ne c$ (a=4.62, b=4.21, c=2.76 {\AA}). 
Interestingly, the lowest-energy B19 TS (Fig.~\ref{Fig2path}f)  and the kinetically--limited intermediate structures in Fig.~\ref{Fig2path}d,e have similar local atomic structure and comparable energies (horizontal dashed line in Figs.~\ref{Fig4} and \ref{Fig3aus}).

{\par} 	For completeness, we consider transformation between B2 and various BCO cells (Fig.~\ref{Fig3aus}), but find no barrier. This barrier\-less path confirms B2 instability. The MEP also differs from the  suggested B2-twin-BCO transformation with a barrier between the twin and BCO  \cite{PRB78p092103y2008}.

{\par}{\bf Austenite-to-Martensite:} The martensitic transition occurs between the low-temperature B19'/BCO martensite and high-temperature austenite, represented by our stable Ni$_{27}$Ti$_{27}$ or Ni$_{54}$Ti$_{54}$ structures. 
Ni$_{27}$Ti$_{27}$ doubled along $a$ (Fig.~\ref{Fig1struc}d) transforms into BCO structure if tripled along $a$ (Fig.~\ref{Fig1struc}a), as shown for the orthorhombic path in Fig.~\ref{Fig2path} (a-d-h).
The MEP from the austenite-to-BCO has a barrier of only $1~me$V/atom above the austenite. 
Such a small barrier agrees with narrow hysteresis (within $40\,$K) in experiment \cite{ActaMat52p3383y2004}. 
We also find other MEPs with energy barriers from $1$ to $3.5~me$V/atom (10 to 40~K) above the austenite. 
In general, such barriers can be strongly affected by non-hydrostatic stress, such as martensitic stress  \cite{ActaMat52p3383y2004}. 

{\par} We find that this transformation proceeds through multiple kinetically--limited intermediate states, characterized by higher density and reduced unit-cell volumes, observed in experiment as a multitude of R-phases \cite{IntJPlasticity39p132y2012,MatSciEngA438p579y2006,PhysicaB324p419y2002}.  
Our DFT energy for these states is $13.2\,$meV/atom above BCO, and lower than austenite. 
These intermediate states %, collectively called an R-phase, 
have similar energies, lattice constants, and densities.
The slope of energy %difference ($E - E_{BCO}$) 
versus MEP coordinate is related to the speed of transformation.
These intermediate states appear along the MEP (Fig.~\ref{Fig3aus}) in the regions labeled R' and B19' in Fig.~\ref{Fig3aus} where the transformation slows down.
Different transformation paths contain similar (but not identical) kinetically limited intermediate states (Fig.~\ref{Fig2path}~d,~e) with local atomic arrangements like a B19 TS in Fig.~\ref{Fig2path}f.
We emphasize that mechanical deformation of martensite and its transformation to austenite upon heating proceeds through similar intermediate structures. 

{\par}Of course, the martensitic transformation happens without diffusion, only by local atomic rearrangements. 
The \emph{ideal} local atomic positions are  ``remembered'' in spite of displacements of atoms (relative to unstable B2 positions) in austenite. 
Austenite has substantial atomic displacive disorder at any temperature (not a classical thermal disorder of a perfectly ordered crystal), but  displacements from ideal B2 do not exceed $1/4$ of the B2 nearest-neighbor distances.  Transformations between these two phases, namely, an easily deformable ``superelastic'' martensite and austenite with arrangement of displacively disordered but ``leashed'' atoms, account for the interesting shape-memory effect observed in NiTi alloys.

%\section{\label{sConclusion}Summary}
{\par} In summary, we addressed the structures and transformation relevant to NiTi shape-memory behavior, both austenite-to-martensite and martensite-to-martensite. 
These transformations include a newly identified austenitic structure, 
intermediate states (collectively called an R-phase), and low-energy deformations of BCO (observed as B19'), as well as the BCO groundstate. 
We also included the (unstable) B2 structure -- historically (but incorrectly) assumed  as the austenite phase in shape-memory studies. 
Together the martensites, austenite and R-phases, and their specific atomic displacements and structural deformations, provide an atomic-scale understanding of the transformations responsible for the NiTi shape-memory effects.
Our results agree with the observed B19' martensitic structure and its super-elasticity. 
The multiple low-energy planar defects, including twins, 
within BCO (needed to form a martensite) agree with the experimental observations \cite{ProgMatSci50n5p511y2005Otsuka}. 
Considering the austenite-to-martensite transformations, we found multiple pathways proceeding through kinetically limited states, which are similar to the lowest-energy B19 transition state (B19' intermediate states) for the BCO-to-BCO transform. Although such intermediate states differ in atomic structure, all of them are similar in energy and density, which agrees with the measured increased density in the R-phase, and the variety of the R-phase structures suggested from experiments.

% Acknowledgments
\acknowledgments
{\bf Acknowledgments: }This work was supported by the U.S. Department of Energy, Office of Science, Basic Energy Sciences, Materials Science and Engineering Division. The research was performed at the Ames Laboratory, which is operated for the U.S. DOE by Iowa State University under contract DE-AC02-07CH11358.

\bibliography{NiTi}
\end{document}